\documentclass[12pt,preprint]{aastex}

\shorttitle{The Black Hole Mass of NGC 4593}
\shortauthors{}

\received{}
\accepted{}

\begin{document}

\title{The Mass of the Black Hole in the Seyfert 1 Galaxy NGC~4593 from Reverberation Mapping}

\author{ Kelly~D.~Denney\altaffilmark{1},
         Misty~C.~Bentz\altaffilmark{1}, 
         Bradley~M.~Peterson\altaffilmark{1}, 
         Richard~W.~Pogge\altaffilmark{1},
         Edward~M.~Cackett\altaffilmark{2}, 
         Matthias~Dietrich\altaffilmark{1},
         Jeffrey~K.~J.~Fogel\altaffilmark{3}, 
         Himel~Ghosh\altaffilmark{1},
         Keith~D.~Horne\altaffilmark{2}, 
         Charles~Kuehn\altaffilmark{1,4},
	 Takeo~Minezaki\altaffilmark{5},
         Christopher~A.~Onken\altaffilmark{1,6},
	 Vladimir~I.~Pronik\altaffilmark{7,8},
         Douglas~O.~Richstone\altaffilmark{3},
	 Sergey~G.~Sergeev\altaffilmark{7,8},
         Marianne~Vestergaard\altaffilmark{9},
         Matthew~G.~Walker\altaffilmark{3}, and
	 Yuzuru~Yoshii\altaffilmark{5,10}}

\altaffiltext{1}{Department of Astronomy, 
		The Ohio State University, 
		140 West 18th Avenue, 
		Columbus, OH 43210; 
		denney, bentz, dietrich, ghosh, peterson, 
                pogge@astronomy.ohio-state.edu}

\altaffiltext{2}{School of Physics and Astronomy, 
		 University of St. Andrews,
		 Fife, KY16 9SS, Scotland, UK;
		 emc14, kdh1@st-and.ac.uk}

\altaffiltext{3}{Department of Astronomy,
		 University of Michigan,
		 Ann Arbor, MI 48109-1090;
		 fogel, dor, mgwalker@umich.edu}

\altaffiltext{4}{Current address:  Physics and Astronomy Department,
		 3270 Biomedical physical Sciences Building,
		 Michigan State University,
		 East Lansing, MI 48824;
		 kuehncha@msu.edu}

\altaffiltext{5}{Institute of Astronomy, 
		 School of Science, 
		 University of Tokyo,
	 	 2-21-1 Osawa, Mitaka, 
		 Tokyo 181-0015, Japan;
		 minezaki, yoshii@ioa.s.u-tokyo.ac.jp}

\altaffiltext{6}{Present address:
		National Research Council Canada, 
		Herzberg Institute of Astrophysics,
		5071 West Saanich Road,
		Victoria, BC  V9E 2E7;
		christopher.onken@nrc-cnrc.gc.ca }

\altaffiltext{7}{Crimean Astrophysical Observatory,
		 p/o Nauchny, 98409 Crimea, Ukraine;
		 sergeev, vpronik@crao.crimea.ua}

\altaffiltext{8}{Isaak Newton Institute of Chile,
	         Crimean Branch, Ukraine}

\altaffiltext{9}{Steward Observatory, 
		The University of Arizona, 
		933 North Cherry Avenue, 
         	Tucson, AZ 85721; 
		mvestergaard@as.arizona.edu}

\altaffiltext{10}{Research Center for the Early Universe, 
		School of Science,
        	University of Tokyo, 
		7-3-1 Hongo, Bunkyo-ku, 
		Tokyo 113-0033, Japan}

\begin{abstract}

We present new observations leading to an improved black hole mass
estimate for the Seyfert $1$ galaxy NGC~4593 as part of a
reverberation-mapping campaign conducted at the MDM Observatory.
Cross-correlation analysis of the H$\beta$ emission-line light curve
with the optical continuum light curve reveals an emission-line time
delay of $\tau_{\rm cent}=3.73 \pm 0.75$ days.  By combining this time
delay with the H$\beta$ line width, we derive a central black hole mass
of $M_{\rm BH} = (9.8 \pm 2.1) \times 10^{6} M_{\odot}$, an improvement
in precision of a factor of several over past results.

\end{abstract}



\section{INTRODUCTION}

Over the past two decades, reverberation mapping
\citep{Blandford82,Peterson93} has emerged as the most widely applicable
method for directly measuring the masses of supermassive black holes
(SMBH) in active galactic nuclei (AGNs).  The appeal of this method
arises from its physical simplicity, despite the complicated physics
within the central regions of these galaxies.  Moreover, unlike other
methods for measuring the masses of SMBHs, it does not require high
angular resolution.  On the other hand, this method is challenging
because it requires a large series of observations, well spaced in time
and taken over a long duration (at least three times the length of the
longest physical timescale of interest).

The reverberation-mapping method depends on continuum flux variations
from the central source, presumably an accretion disk around a SMBH.
Continuum photons are absorbed and then re-emitted as line photons in the
broad line-emitting region (BLR).  Variations in the broad-line flux
have a time lag relative to the continuum, corresponding to the
light-travel time across the BLR.  Through measurements of this time
lag, the size of the region can be determined, which then leads, through
virial arguments and a measurement of the broad-line width, to the black
hole mass, $M_{\rm BH}$.

To date, black hole masses have been determined for nearly 40 AGNs using
reverberation mapping \citep[see the compilation by][]{Peterson04}.
While this demonstrates how far the field has progressed,
\citet{Peterson04} make it clear that substantial work is still needed
to decrease uncertainties in time lag, and thus $M_{\rm BH}$,
measurements in order to better test and calibrate the various black
hole mass-dependent scaling relationships, such as the $M_{\rm
BH}-\sigma_{\star}$ relationship, where $\sigma_{\star}$ is the bulge
velocity dispersion.  This need for better reverberation-based mass
estimates is the driving force behind the campaign we describe below.

One requirement for a successful reverberation mapping program is good
temporal sampling \citep{Horne04}.  The major goal of the current
observing program is to improve the sampling rate for
reverberation-mapped AGNs that were undersampled in previous campaigns.
The mass of the black hole in NGC~4593 was previously measured by
reverberation mapping \citep{Dietrich94,Onken03}, but the H$\beta$ lag
determination of $\tau_{\rm cent}=3.1$ days was unfortunately much
smaller than the average interval between observations of 15.8 days.
This resulted in uncertainties ($+7.1,-5.1$ days) much larger than the
time delay itself, and as such, were consistent with no lag at all.  By
revisiting this object with our program, we were able to achieve the
sampling necessary to drive down the previous high uncertainties in the
time lag and black hole mass.  Here, we present new lag and $M_{\rm BH}$
determinations from reverberation mapping of NGC~4593.  The
observational uncertainties from this campaign represent a factor of
several improvement over previous results.

\section{OBSERVATIONS AND DATA ANALYSIS}

Observations of NGC~4593 were obtained as part of a large reverberation
mapping campaign undertaken in early 2005 on the 1.3m McGraw-Hill
telescope of the MDM Observatory on Kitt Peak in Arizona.  Details
beyond the descriptions below, for our spectral and photometric
observations and data reduction, are given by \citet{Bentz06b}.
NGC~4593 is one of three AGNs from which we were able to observe
sufficient variability within the time frame of our program to warrant a
reverberation analysis.  It was first recognized as a host to a type~1
AGN by \citet{Lewis78} as part of the Michigan-Tololo Curtis Schmidt
survey for extragalactic emission-line objects
\citep[see][]{MacAlpine79}.  NGC~4593 is a type (R)SB(rs)b spiral galaxy
at a redshift of $z = 0.0090$.

\subsection{Spectroscopy}

{\bf MDM Observations}: We obtained spectra of the central region of
NGC~4593 at 24 epochs between 2005 February and April with the Boller
and Chivens CCD spectrograph on the 1.3-meter telescope at MDM
Observatory, at a spatial scale of 0\farcs75 per pixel.  For our
campaign, we used a grating of 350 grooves/mm, corresponding to a
dispersion of 1.33~\AA/pixel.  The spectral coverage was from $\sim
4300-5700$ \AA, centered on H$\beta\,\lambda 4861$ and the [O\,{\sc
iii}]\,$\lambda \lambda 4959, 5007$ lines, resulting in a spectral
resolution of 7.6 \AA\ across this range.  The slit width was set to
5\farcs0 projected on the sky, with a position angle of $90 \degr$.
Figure 1 shows the mean and rms spectrum of NGC~4593, made from the
complete set of MDM observations.

{\bf CrAO Observations}:  We also acquired spectra of the nuclear region
of NGC~4593 from the Nasmith spectrograph with the Astro-550 $580
\times 520$ pixel CCD \citep{Berezin91} on the 2.6-m Shajn telescope of
the Crimean Astrophysical Observatory (CrAO).  For these observations a
3\farcs0 slit was utilized, with a $90 \degr$ position angle.  Spectral
wavelength coverage for this data set was from $\sim 4300-5600$ \AA,
with a dispersion of 2.0 \AA/pix and a spectral resolution of 8.2 \AA.

\subsection{Photometry}

{\bf MAGNUM Observations}: In addition to spectral observations, we also
obtained $V$-band photometry from the 2.0-m Multicolor Active Galactic
NUclei Monitoring (MAGNUM) telescope at the Haleakala Observatories in
Hawaii, imaged with the multicolor imaging photometer (MIP) as described
by \citet{Kobayashi98a,Kobayashi98b}, \citet{Yoshii02}, and
\citet{Yoshii03}.  Photometric reduction of NGC~4593 was similar to that
described for other sources by \citet{Minezaki04} and
\citet{Suganuma06}, except the host-galaxy contribution to the flux
within the aperture was not subtracted and the filter color term was not
corrected because these photometric data were later scaled to the MDM
and CrAO continuum light curve (as described below).  Also, minor
corrections (of order 0.01 mag or less) due to the seeing dependence of
the host-galaxy flux were ignored.

\subsection{Light Curves}

Following reduction of the data, light curves were created for the
subsequent cross-correlation analysis.  The spectral fluxes of the MDM
observations were calibrated based on the [O\,{\sc iii}]\,$\lambda 5007$
line flux between observed frame wavelengths of $5022-5062$ \AA\ in the
mean spectrum.  Individual spectra were scaled to this reference
spectrum using software that employs a $\chi^{2}$ goodness of fit
estimator \citep{vanGroningen92}.  Because few MDM observations were
taken under photometric conditions, our absolute [O\,{\sc
iii}]\,$\lambda 5007$ flux calibrations are not reliable.  Therefore,
the final light curves are calibrated to the [O\,{\sc iii}]\,$\lambda
5007$ flux given by \citet{Dietrich94}.  Light curves from the CrAO
spectra were made following a similar method, only the final light curve
flux measurements were not calibrated to the \citet{Dietrich94} value
because this data set was later scaled to the MDM data set, as described
below.  Continuum and line flux measurements were obtained by first
fitting the continuum on either side of the H$\beta$ line, between
wavelength ranges $4795-4815$ \AA\ and $5120-5170$ \AA\ in the observed
frame.  The H$\beta$ flux was calculated by integrating above the
continuum over the wavelength range $4825-4963$ \AA.  The continuum flux
density was then taken to be the average over $5120-5170$ \AA.  Although
at least some contamination from Fe\,{\sc ii} emission is unavoidable
when measuring the continuum in the optical, we chose this continuum
region because it is the cleanest window, with respect to this
contamination, within the spectral coverage of our observations. In
addition, \citet{Vestergaard05} demonstrate that although Fe\,{\sc ii}
emission is variable, the amplitude of variability is generally low.  We
do not expect Fe\,{\sc ii} contamination to significantly affect our
measurements.

In order to intercalibrate the data from various sources (i.e. place
them on a consistent flux scale), corrections to the measured line and
continuum flux values are necessary because of systematic differences,
mostly attributable to seeing and aperture effects
\citep[see][]{Peterson91}.  Given data sets from two sources, flux
differences were determined by comparing all possible pairs of
simultaneous observations between the two data sets.  Here, we relaxed
the meaning of `simultaneous' to include pairs separated by at most 2.0
days so as to increase the number of pairs contributing to each
comparison.  However, this relaxed assumption does not produce
significantly larger (i.e. less than $1.0\%$) uncertainties, implying
that there is no evidence for intrinsic variability on such short
timescales.  Based on the average difference in flux between these
closely spaced pairs, a single multiplicative point-source correction
factor, $\varphi_{\rm CrAO} = 0.89$, was applied to all points in both
the emission-line and continuum light curves of the secondary data set
(CrAO) to scale it to the primary set (MDM), following the methods of
\citet{Peterson91}.  Physically, this scaling factor accounts for
differences in the amount of [O\,{\sc iii}]\,$\lambda 5007$ flux
measured due to different aperture geometries and seeing between the two
data sets.  Seeing can also effect the internal calibrations of the
continuum and H$\beta$ emission-line flux measurements, based on
$F$([O\,{\sc iii}]\,$\lambda 5007$), which is assumed to be constant.
However, given the large slit widths used in the current MDM and CrAO
observations (5\farcs0 and 3\farcs0, respectively) as well as the
compact (no larger than 1\farcs7) emission region of [O\,{\sc iii}] in
NGC~4593 \citep{Schmitt03}, seeing is an insignificant effect in our
internal calibrations.  Further tests of the effect of seeing on our
internal flux calibrations will be addressed in \S 3.

In addition to this multiplicative correction, an additive correction
factor, $G_{\rm CrAO} = -1.60 \times 10^{-15}$ erg s$^{-1}$ cm$^{-2}$
\AA$^{-1}$, was also made to the continuum flux, accounting for the
starlight contribution in the different apertures of the various data
sets.  These correction methods were employed to scale both the
continuum and H$\beta$ light curves from the CrAO observations to the
MDM data, since the MDM data formed the largest, individual set, and
thus served as the base onto which to tie the others.  For the same
reason, the photometric, $V$-band observations from the MAGNUM data were
then scaled to the newly combined MDM and CrAO continuum light curve,
with additive correction factor $G_{\rm MAGNUM} = 0.65 \times 10^{-15}$
erg s$^{-1}$ cm$^{-2}$ \AA$^{-1}$.  The continuum light curve was
finally corrected for host galaxy starlight, where, for this slit
geometry, the contribution is measured to be
$F_{gal}$(5100\AA)$=(1.11^{+0.10}_{-0.11}) \times 10^{-14}$ erg s$^{-1}$
cm$^{-2}$ \AA$^{-1}$ from a recent {\it Hubble Space Telescope} image
taken with the Advanced Camera for Surveys High Resolution Camera.  The
methods by which this flux contribution was determined are described by
\citet{Bentz06a}, and the complete results specific to this object will
appear in a future paper.  The full set of observations used in
constructing these light curves can be found in Table 1, and Figure~2
shows the corresponding light curves.

\section{TIME SERIES ANALYSIS}

For the time series analysis, the light curves from Figure~2 were
modified by first merging the three sets of observations by binning data
points more closely spaced than 0.5 day through the use of a
variance-weighted average.  Cross-correlation analysis was restricted to
the subset of the light curves where both continuum and emission-line
measurements are available, i.e. JD2453430 -- JD2453472.  In addition,
two observations from the CrAO data set, JD2453463.4 and JD2453465.9,
show anomalously high continuum flux for reasons that could not be
determined.  Conducting the cross-correlation analysis including these
points resulted in lag determinations consistent with removing the
points, although the uncertainties in the lags were higher.  Since this
time period was otherwise well sampled, none of the other observations
show a similar peak in continuum flux, and because the results from
exclusion are consistent with inclusion, we omit these two continuum
points from the final light curve.

Figure 3 shows the modified light curves that were used for the cross
correlation analysis.  The statistical parameters describing these final
light curves are shown in Table~2.  Column (1) gives the spectral
feature, and Column (2) shows the number of data points in each light
curve.  Columns (3) and (4) are mean and median sampling intervals,
respectively, between data points. The mean flux with standard deviation
is given in column (5), while column (6) shows the mean fractional
error, based on comparison between closely spaced observations. Column
(7) gives the excess variance, calculated as

\begin{equation}
F_{\rm var} = \frac{\sqrt{\sigma^2 - \delta^2}}{\langle f \rangle}
\end{equation}

\noindent where $\sigma^2$ is the variance of the observed fluxes, 
$\delta^2$ is their mean square uncertainty, and $\langle f \rangle$ is
the mean of the observed fluxes.  Finally, column (8) is the ratio of
the maximum to minimum flux in the light curves.

We conducted the time-series analysis by cross correlating the continuum
light curve with the H$\beta$ light curve using two methods designed for
unevenly spaced observations.  The primary method uses an interpolation
scheme that averages two results: first, cross correlating an
interpolated continuum light curve with the original emission-line light
curve, and second, cross correlating the original continuum with an
interpolated emission-line light curve \citep{Gaskell87}.  We employed
an interpolation interval of 0.5 day (equal to half of the median time
span between observations), chosen to create equal spacing on a scale
smaller than that of the actual data, yet large enough to prevent the
introduction of artifacts due to the correlation of adjacent
interpolated flux values.  The correlation coefficient, $r$, is computed
from pairs of points from each light curve matched based on different
lags being imposed upon the trailing curve.  For a given time lag
realization, unpaired points at the beginning or end of the time series
are excluded from the correlation analysis.  By calculating $r$ for
multiple potential lag times, a cross-correlation function (CCF) is
constructed, shown in Figure~4, which gives the correlation coefficient
at each of these different lag times, $\tau$.  From the CCF, we can
characterize the time delay through two parameters.  The peak lag,
$\tau_{\rm peak}$, is simply taken as the lag time which produces the
highest correlation coefficient, $r_{\rm max}$, whereas the centroid
lag, $\tau_{\rm cent}$, is the centroid of the CCF, computed from values
with $r\geq 0.8r_{\rm max}$.

The uncertainties in the time lag measurements were computed via
model-independent Monte-Carlo simulations and based on the bootstrap
method referred to as FR/RSS (for flux redistribution/random subset
selection), described by \citet{Peterson98} and which utilized the
modifications of \citet{Peterson04}. A large number of realizations of
this method (e.g. 10,000 in this study) are employed to build up two
distributions of CCFs: the cross correlation centroid distribution
(CCCD) and the cross correlation peak distribution (CCPD).  The values
$\tau_{\rm cent}$ and $\tau_{\rm peak}$ are the means of these
respective distributions.  The uncertainties are calculated such that
15.87\% of the realizations yield values larger than the mean plus the
upper error, and 15.87\% yield values smaller than the mean minus the
lower error; these would correspond to the $1\sigma$ errors in the case
of a Gaussian distribution.  Table~3 gives the calculated values of
$\tau_{\rm cent}$ and $\tau_{\rm peak}$ for NGC~4593, after being
corrected for time dilation.

The second cross correlation method we employ creates a discrete
correlation function (DCF), described by \citet{Edelson88}, including
modifications by \citet{White94}. Shown in Figure~4, the DCF is created by
determining correlations at different lag times between the H$\beta$ and
continuum light curves, just as in the interpolation method, but through
time binning of data rather than interpolation.  Under-sampled data or
data with large time gaps could lead to spurious lag determinations
through the interpolation method, in which case there are advantages to
using the DCF method.  For example, only the actual data are used in the
analysis and due to the binning, statistical uncertainties can be
assigned to the correlation coefficient for each bin
\citep[see][]{White94}.  For this analysis, a time bin of 2.0 days was
adopted.  One can see from Figure~4 that the DCF agrees well with the
CCF from the interpolation method showing that interpolation is not
introducing artifacts.  Figure 4 also shows the auto-correlation
function (ACF), which is computed by cross correlating the continuum
with itself. The ACF is interesting because the CCF is the convolution
of this distribution (ACF) and the transfer function, which describes
the emission-line response to the continuum variations and is the
quantity reverberation mapping is fundamentally attempting to recover
\citep[see][]{Peterson93}.

The cross-correlation functions in Figure 4 also demonstrate that our
continuum and H$\beta$ emission-line internal flux calibration to the
[O\,{\sc iii}]\,$\lambda 5007$ emission line are not adversely affected
by nightly changes in seeing.  Because calibration errors due to seeing
would result in correlated errors, with both the continuum and the line
flux affected in the same direction, if variable seeing is a significant
effect, it would lead to a spike in the cross-correlation function at a
lag of zero days.  Figure 4 clearly shows no such feature in either the
CCF or DCF for NGC~4593.  \citet{Bentz06b} report on results for
NGC~4151 from this same observing campaign and do not see any evidence
for this zero lag spike either.  The only object from this campaign in
which a zero-lag spike is detected in the cross-correlation function is
NGC~5548 \citep{Bentz06c}, which was in a very low-flux state at the
time.  In this case, the amplitude of variability was so low that the
weak correlated-error signature was above the detection threshold.

To investigate the possibility of detecting bulk motions within the BLR,
we then broke the H$\beta$ broad emission-line into velocity bins and
performed the cross correlation anaylsis between these individual bins.
First, the H$\beta$ line was divided in half based on the peak in the
mean spectrum (see Fig. 1), with the blue side defined by wavelengths
4825 -- 4903 \AA, and the red side defined by wavelengths 4903 -- 4963
\AA.  The flux was determined by integrating the line flux above the
same continuum fit as described above.  The lags determined through this
cross correlation analysis are $\tau_{\rm cent} = 0.5 \pm 1.0$ day and
$\tau_{\rm peak} = 0.7^{+1.0}_{-1.5}$ day, with variations in the blue
side slightly lagging behind those in the red side.  Second, the line
was broken into a `core' and `wings' such that the integrated flux in
the core was equal to the summed flux in both red and blue wings.  These
regions were defined from the mean spectrum such that the blue wing
constituted the integrated flux between wavelengths 4825 -- 4881.8 \AA,
the core between 4881.8 -- 4924.2, and the red wing between 4924.2 --
4963 \AA.  The lags determined for this scenario are $\tau_{\rm cent} =
0.5 \pm 0.7$ day and $\tau_{\rm peak} = 0.1 \pm 1.0$ day, with
variations in the wings leading the core as would be expected.  The CCFs
from both of these analyses are shown in Figure 5.  Unfortunately, any
signature of inflow/outflow or bulk Keplarian motions was too weak to be
detected with confidence through this investigation, as the lags
measured for both scenarios were consistent with zero.

\section{BLACK HOLE MASS}

Assuming that the motions within the line-emitting region are
gravitationally dominated so that the virial theorem can be utilized,
the mass of the black hole can be defined such that

\begin{equation}
M_{\rm BH} = \frac{f c \tau (\Delta V)^2}{G},
\end{equation}

\noindent where $\tau$ is the measured emission-line time delay and 
$\Delta V$ is the emission-line width, which can be characterized by
either the full width at half maximum (FWHM) of the broad emission-line
or by the emission-line dispersion, $\sigma_{\rm line}$.  The
dimensionless factor $f$ depends on the structure, kinematics, and
inclination of the BLR and is of order unity.  By normalizing the
reverberation-based black hole masses to the $M_{\rm BH} -
\sigma_{\star}$ relationship of quiescent galaxies, \citet{Onken04} have
shown that $f$ has an average value of 5.5, when the emission-line
dispersion, or second moment of the profile, is used (as opposed to the
FWHM\footnote{See \citet{Collin06} for a discussion of systematic
differences between the use of the FWHM and the line dispersion in
calculating the virial product, and thus $M_{\rm BH}$.}) for $\Delta V$.

After first removing the narrow line component of the H$\beta$ emission
line, the FWHM and the line dispersion were measured from both the mean
and the rms spectra of NGC~4593.  \citet{Peterson04} describe in detail
the bootstrap method used to determine these quantities and their
respective uncertainties. The measured values for the FWHM and
$\sigma_{\rm line}$ are given in Table 3.
	
To calculate the black hole mass for NGC~4593, we use the centroid lag,
$\tau_{\rm cent}$, for the time delay, $\tau$, and the line dispersion,
$\sigma_{\rm line}$, of the H$\beta$ emission line, measured from the
rms spectrum, for the emission-line width, $\Delta V$.
\citet{Peterson04} argue that this combination of parameters gives
the most reliable black hole mass determinations, based on virial
arguments and fits between all possible combinations of parameters.  The
black hole mass calculated for NGC~4593 from this work is $M_{\rm BH} =
(9.8 \pm 2.1)\times 10^{6}M_{\odot}$, where the uncertainties quoted
include statistical and observational considerations but not an
intrinsic uncerainty in the method, no larger than a factor of 2-3, that
accounts for unknowns such as inclination.

Our result is consistent with the mass estimate based on a previous
reverberation program.  \citet{Onken03} reanalyzed observations of
NGC~4593 by the ``Lovers of Active Galaxies (LAG)'' consortium
\citep{Dietrich94} and determined a black hole mass based on H$\beta$ to
be $M_{\rm BH} = (5.4^{+9.4}_{-7.0}) \times 10^{6}M_{\odot}$, where we
have used the \citet{Onken04} calibration \citep{Peterson04}.  The
improved time sampling of the new reverberation mapping observations has
significantly decreased the uncertainties in the black hole mass
estimate for this object.

\section{DISCUSSION AND CONCLUSION}

Results presented here for the emission-line lag and black hole mass of
NGC~4593 of $\tau_{\rm cent}=3.73 \pm 0.75$ days and $M_{\rm BH} = (9.8
\pm 2.1) \times 10^{6} M_{\odot}$, for these respective quantities,
represent a factor of several improvement over past measurements.
Previous measurements for this object from LAG data exhibited much
larger uncertainties due in part to the average sampling interval, which
due mostly to bad weather, was much larger (by about a factor of 4) than
the emission-line lag of this object.  In addition, for rather short
observing campaigns, some of the success of reverberation mapping comes
from serendipitously observing a large variability event during the
campaign.  The best results come from observing a large increase and
then decrease (or vice versa) in the flux, which we observed in this
campaign but was unfortunately not seen to the same degree in the
previous campaign of NGC~4593.  This represents yet another example
exhibiting the need for longer observing campaigns.

The results reported here were obtained as part of a larger
spectroscopic monitoring campaign whose primary goal was to improve the
emission-line lag and black hole mass measurements for some of the
nearest, apparently brightest AGNs.  The proximity of these AGNs makes
them especially important not only as calibrators of the AGN $M_{\rm
BH}$--$\sigma_*$ and BLR radius--luminosity relationships
\citep{Bentz06a}, but also as candidates for measurement of their black
hole masses through other means that depend on high angular
resolution. Key elements of this program were (1) scheduled daily
observations of each of these relatively low-luminosity AGNs, (2)
supporting observations at multiple sites to mitigate the effects of
gaps in the time series caused by weather, and (3) high-quality
homogeneous data that permitted relative spectrophotometric flux
calibration at better than the 2\% level. Our 42-night program yielded
improved H$\beta$ lags and black hole masses with a factor of several
improvement in precision relative to previous campaigns for three of the
six AGNs in our sample, NGC 4151 \citep{Bentz06b}, NGC 5548, which was
observed in the lowest-luminosity state yet recorded (Bentz et al., in
preparation), and NGC 4593, as described here.  Three other AGNs in our
program, NGC 3227, NGC 3516, and NGC 4051 were insufficiently variable
during this relatively brief campaign, underscoring the point that
longer duration programs are necessary for detection of variability
signatures that are favorable for reverberation analysis.

\acknowledgements
We acknowledge support for this work by the National Science Foundation
though grant AST-0205964 to The Ohio State University and the Civilian
Research and Development Foundation through grant UP1-2549-CR-03.
K.~D. is supported by a GK-12 Fellowship of the National Science
Foundation.  M.~B. is supported by a Graduate Fellowship of the National
Science Foundation.  E.~C. gratefully acknowledges support from PPARC.
M.~V. acknowledges financial support from NSF grant AST-0307384 to the
University of Arizona.  This research has made use of the NASA/IPAC
Extragalactic Database (NED) which is operated by the Jet Propulsion
Laboratory, California Institute of Technology, under contract with the
National Aeronautics and Space Administration.


\clearpage


\begin{deluxetable}{cccc}
\tablecolumns{4}
\tablewidth{0pt}
\tablecaption{Continuum and H$\beta$ Fluxes for NGC~4593}
\tablehead{
\colhead{JD} &
\colhead{F$_{\lambda}$ (5100 \AA)} &
\colhead{H$\beta$ $\lambda 4861$} &
\colhead{Data Set}\\
\colhead{(-2450000)} &
\colhead{($10^{-15}$ erg s$^{-1}$ cm$^{-2}$ \AA$^{-1}$)} &
\colhead{($10^{-13}$ erg s$^{-1}$ cm$^{-2}$)}}

\startdata

3391.986 & 10.69 $\pm$ 0.10   & \nodata	     & MAGNUM \\
3411.926 &  9.23 $\pm$ 0.10   & \nodata	     & MAGNUM \\
3419.121 &  9.86 $\pm$ 0.05   & \nodata	     & MAGNUM \\
3430.844 &  8.94 $\pm$ 0.18   & 4.59 $\pm$  0.09 & MDM \\
3430.962 &  9.78 $\pm$ 0.08   & \nodata	     & MAGNUM \\ 
3431.832 &  8.75 $\pm$ 0.18   & 4.52 $\pm$  0.09 & MDM \\
3433.816 &  8.52 $\pm$ 0.17   & 4.61 $\pm$  0.09 & MDM \\
3437.499 &  9.68 $\pm$ 0.32   & 4.67 $\pm$  0.10 & CrAO \\
3437.926 &  9.39 $\pm$ 0.19   & 4.77 $\pm$  0.10 & MDM \\
3438.073 & 10.37 $\pm$ 0.04   & \nodata 	     & MAGNUM \\ 
3438.805 &  9.81 $\pm$ 0.20   & 5.03 $\pm$  0.10 & MDM \\
3439.789 &  9.60 $\pm$ 0.19   & 5.00 $\pm$  0.10 & MDM \\
3440.824 &  9.51 $\pm$ 0.19   & 5.08 $\pm$  0.10 & MDM \\
3441.848 &  9.24 $\pm$ 0.19   & 5.12 $\pm$  0.10 & MDM \\
3442.820 &  8.92 $\pm$ 0.18   & 5.34 $\pm$  0.11 & MDM \\
3443.820 &  9.05 $\pm$ 0.18   & 5.23 $\pm$  0.11 & MDM \\
3444.450 &  8.75 $\pm$ 0.29   & 5.08 $\pm$  0.11 & CrAO \\
3445.456 &  7.99 $\pm$ 0.26   & 5.07 $\pm$  0.11 & CrAO \\
3445.832 &  8.80 $\pm$ 0.18   & 5.01 $\pm$  0.10 & MDM \\
3446.448 &  8.64 $\pm$ 0.29   & 4.91 $\pm$  0.10 & CrAO \\
3446.816 &  8.63 $\pm$ 0.17   & 4.97 $\pm$  0.10 & MDM \\
3450.832 &  8.19 $\pm$ 0.16   & 4.53 $\pm$  0.09 & MDM \\
3451.840 &  8.38 $\pm$ 0.17   & 4.36 $\pm$  0.09 & MDM \\
3452.793 &  8.66 $\pm$ 0.17   & 4.27 $\pm$  0.09 & MDM \\
3459.848 &  8.99 $\pm$ 0.18   & 4.56 $\pm$  0.09 & MDM \\
3460.809 &  8.71 $\pm$ 0.17   & 4.58 $\pm$  0.09 & MDM \\
3461.812 &  9.00 $\pm$ 0.18   & 4.56 $\pm$  0.09 & MDM \\ 
3462.848 &  9.00 $\pm$ 0.18   & 4.63 $\pm$  0.09 & MDM \\
3463.425 & 10.23 $\pm$ 0.34   & 4.78 $\pm$  0.10 & CrAO\tablenotemark{a} \\ 
3464.426 &  9.58 $\pm$ 0.32   & 4.76 $\pm$  0.10 & CrAO\tablenotemark{a} \\ 
3465.746 &  9.19 $\pm$ 0.18   & 4.80 $\pm$  0.10 & MDM \\
3465.890 &  9.93 $\pm$ 0.15   & \nodata 	     & MAGNUM \\
3467.844 &  9.19 $\pm$ 0.18   & 4.70 $\pm$  0.09 & MDM \\
3469.355 &  9.33 $\pm$ 0.31   & 4.87 $\pm$  0.10 & CrAO \\
3469.785 &  9.41 $\pm$ 0.19   & 4.80 $\pm$  0.10 & MDM \\
3470.364 &  9.34 $\pm$ 0.31   & 5.05 $\pm$  0.11 & CrAO \\
3470.852 &  9.54 $\pm$ 0.19   & 5.00 $\pm$  0.10 & MDM \\
3471.848 &  9.29 $\pm$ 0.19   & 4.90 $\pm$  0.10 & MDM \\
3478.823 &  9.60 $\pm$ 0.07   & \nodata 	     & MAGNUM \\
3485.007 &  9.70 $\pm$ 0.07   & \nodata 	     & MAGNUM \\
3508.910 &  9.91 $\pm$ 0.05   & \nodata 	     & MAGNUM \\
3527.847 &  9.99 $\pm$ 0.03   & \nodata 	     & MAGNUM \\
3539.818 & 10.61 $\pm$ 0.12   & \nodata 	     & MAGNUM  \\
3556.769 & 10.46 $\pm$ 0.06   & \nodata 	     & MAGNUM \\
3572.766 &  8.44 $\pm$ 0.12   & \nodata 	     & MAGNUM \\
3579.766 &  9.35 $\pm$ 0.25   & \nodata	     & MAGNUM \\

\enddata

\tablenotetext{a}{Continuum point omitted from final light curve.  See 
\S 3.}

\end{deluxetable}

\clearpage

\begin{deluxetable}{cccccccc}
\tablecolumns{8}
\tablewidth{0pt}
\tablecaption{Light Curve Statistics}
\tablehead{
\colhead{ }&\colhead{ }&\multicolumn{2}{c}{Sampling}&\colhead{ }&
\colhead{Mean}&\colhead{ }&\colhead{ }\\
\colhead{Time}&\colhead{ }&\multicolumn{2}{c}{Interval(days)}&\colhead{Mean}&
\colhead{Fractional}&\colhead{ }&\colhead{ }\\
\colhead{Series}&\colhead{$N$}&\colhead{$\langle T \rangle$}&
\colhead{$T_{\rm median}$}&\colhead{Flux\tablenotemark{a}}&\colhead{Error}&
\colhead{$F_{\rm var}$}&\colhead{$R_{\rm max}$}\\
\colhead{(1)}&\colhead{(2)}&\colhead{(3)}&\colhead{(4)}&\colhead{(5)}&
\colhead{(6)}&\colhead{(7)}&\colhead{(8)}}
\startdata

$5100$ \AA & $25$ & $1.7$ & $1.0$ & $4.8 \pm 0.7$ & $0.06$ & $0.14$ & $1.87 \pm 0.19$\\
 H$\beta$ & $27$ & $1.6$ & $1.0$ & $8.4 \pm 0.5$ & $0.02$ & $0.05$ & $1.25 \pm 0.04$\\


\enddata
\tablenotetext{a}{Same flux units as Table 1 for $5100$ \AA\ continuum and H$\beta$, respectively.}
\end{deluxetable}

\clearpage

\begin{deluxetable}{cc}
\tablecolumns{2}
\tablewidth{0pt}
\tablecaption{Reverberation Results}
\tablehead{
\colhead{Parameter}&{Value}\\
\colhead{(1)}&{(2)}}
\startdata
$\tau_{\rm cent}$ & $3.73 \pm 0.75$ days\\
$\tau_{\rm peak}$ & $3.4^{+0.5}_{-1.0}$ days\\
$\sigma_{\rm line} (\rm mean)$ & $1790 \pm 3$ km s$^{-1}$\\
FWHM (mean)& $5143 \pm 16$ km s$^{-1}$\\
$\sigma_{\rm line} (\rm rms)$ & $1561 \pm 55$ km s$^{-1}$\\
FWHM (rms)& $4141 \pm 416$ km s$^{-1}$\\
$M_{\rm BH}$ & $(9.8 \pm 2.1) \times 10^{6}M_{\odot}$\\
\enddata
\end{deluxetable}

\clearpage


\begin{figure}
\figurenum{1}
\epsscale{1}
\plotone{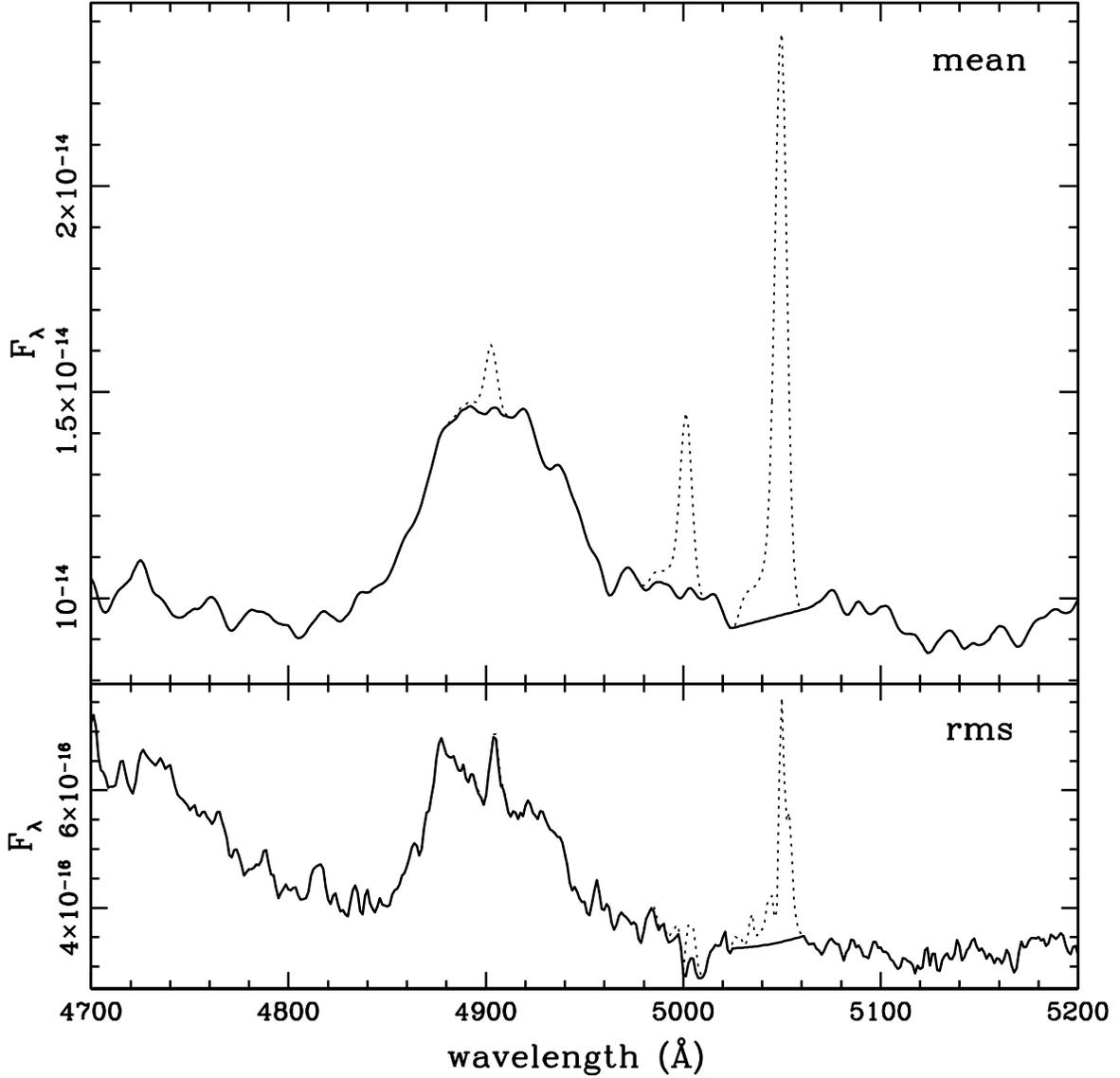}
\caption{Mean and RMS spectrum of NGC~4593 from MDM observations.  The solid line shows the spectrum with the narrow-line components of H$\beta$, [O\,{\sc iii}]\,$\lambda 4959$, and [O\,{\sc iii}]\,$\lambda 5007$ removed.  The dotted line shows where these narrow-line components contributed to the spectrum before they were removed. The large increase in rms flux shortward of 4800 \AA\ is due to variations in the broad He\,{\sc ii}\,$\lambda 4686$ emission line.}
\end{figure}

\begin{figure}
\figurenum{2}
\epsscale{1}
\plotone{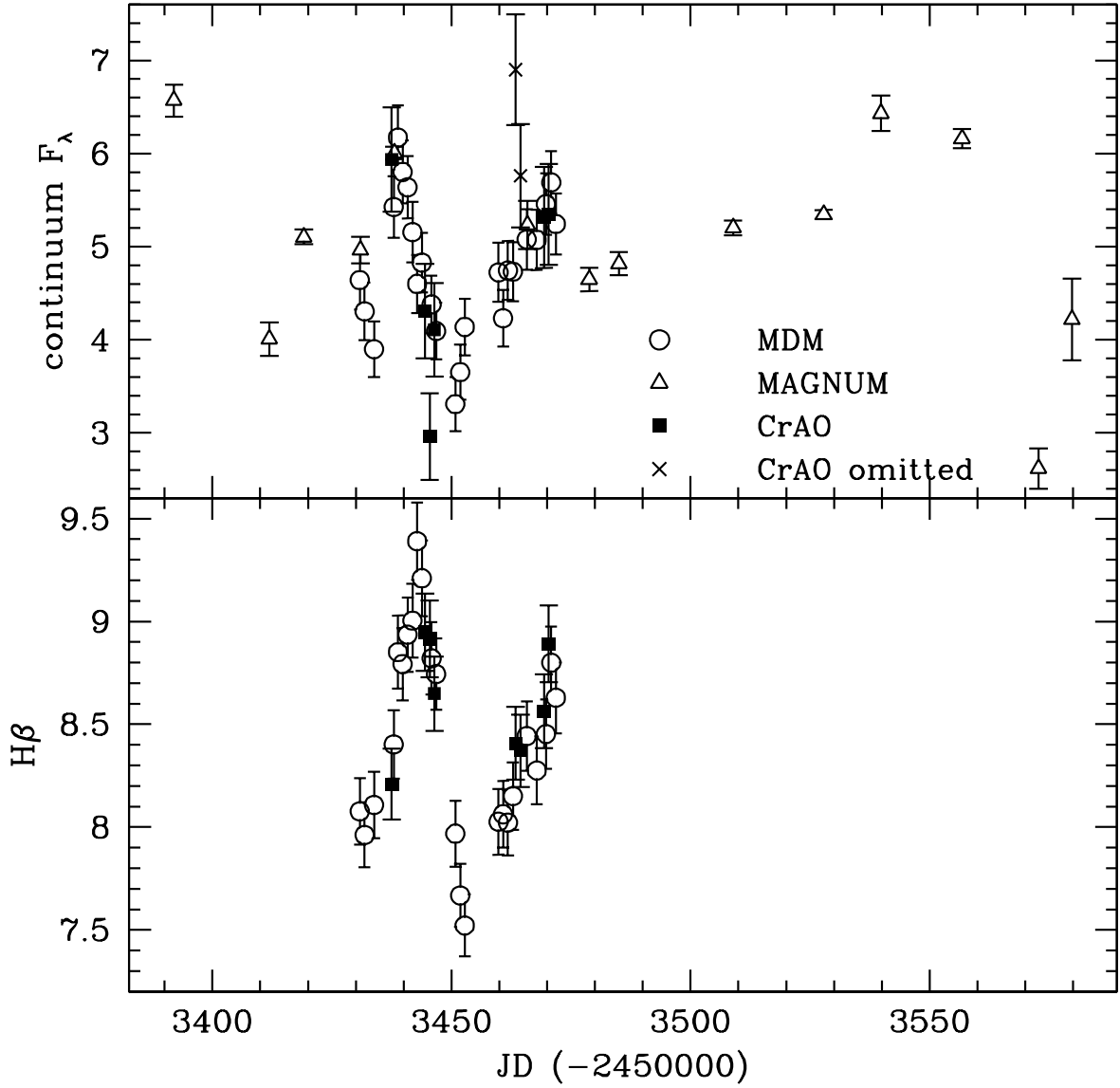}
\caption{Light curve showing complete data sets from all three sources.  The top panel shows the 5100 \AA\ continuum flux in units of $10^{-15}$ erg s$^{-1}$ cm$^{-2}$ \AA$^{-1}$, while the bottom is the H$\beta$ $\lambda$4861 line flux in units of $10^{-13}$ erg s$^{-1}$ cm$^{-2}$.  The crosses show the two continuum points that were later omitted from the final light curve (see \S 3).}
\end{figure}

\begin{figure}
\figurenum{3}
\epsscale{1}
\plotone{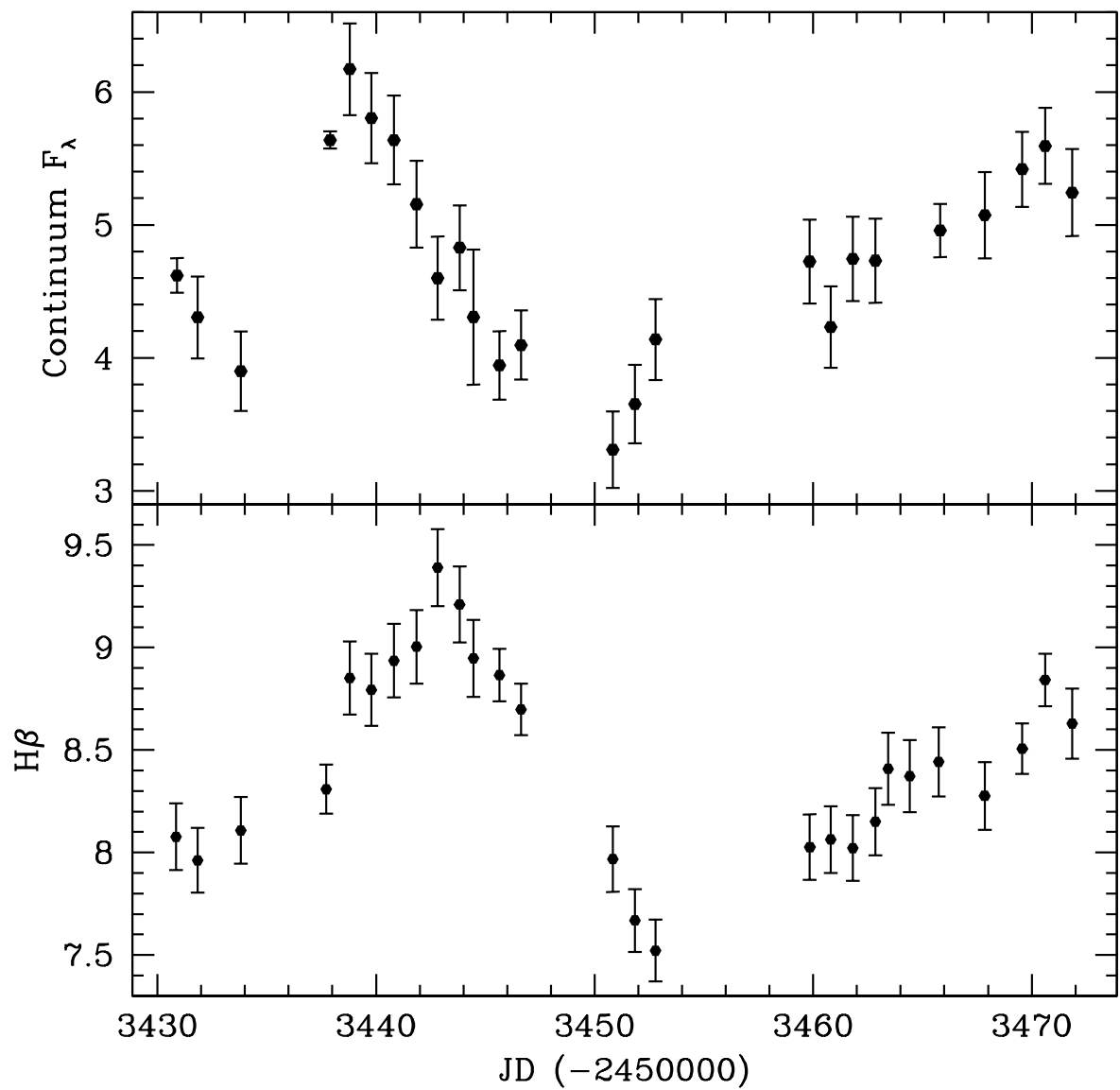}
\caption{Light curve showing subset of data constrained to the time frame of the MDM observations that was used for time series analysis.  Observations within 0.5 day of each other have been averaged together. The top panel shows the 5100 \AA\ continuum in units of $10^{-15}$ erg s$^{-1}$ cm$^{-2}$ \AA$^{-1}$, while the bottom is the H$\beta$ $\lambda$~4861 line flux in units of $10^{-13}$ erg s$^{-1}$ cm$^{-2}$. }
\end{figure}

\begin{figure}
\figurenum{4}
\epsscale{1}
\plotone{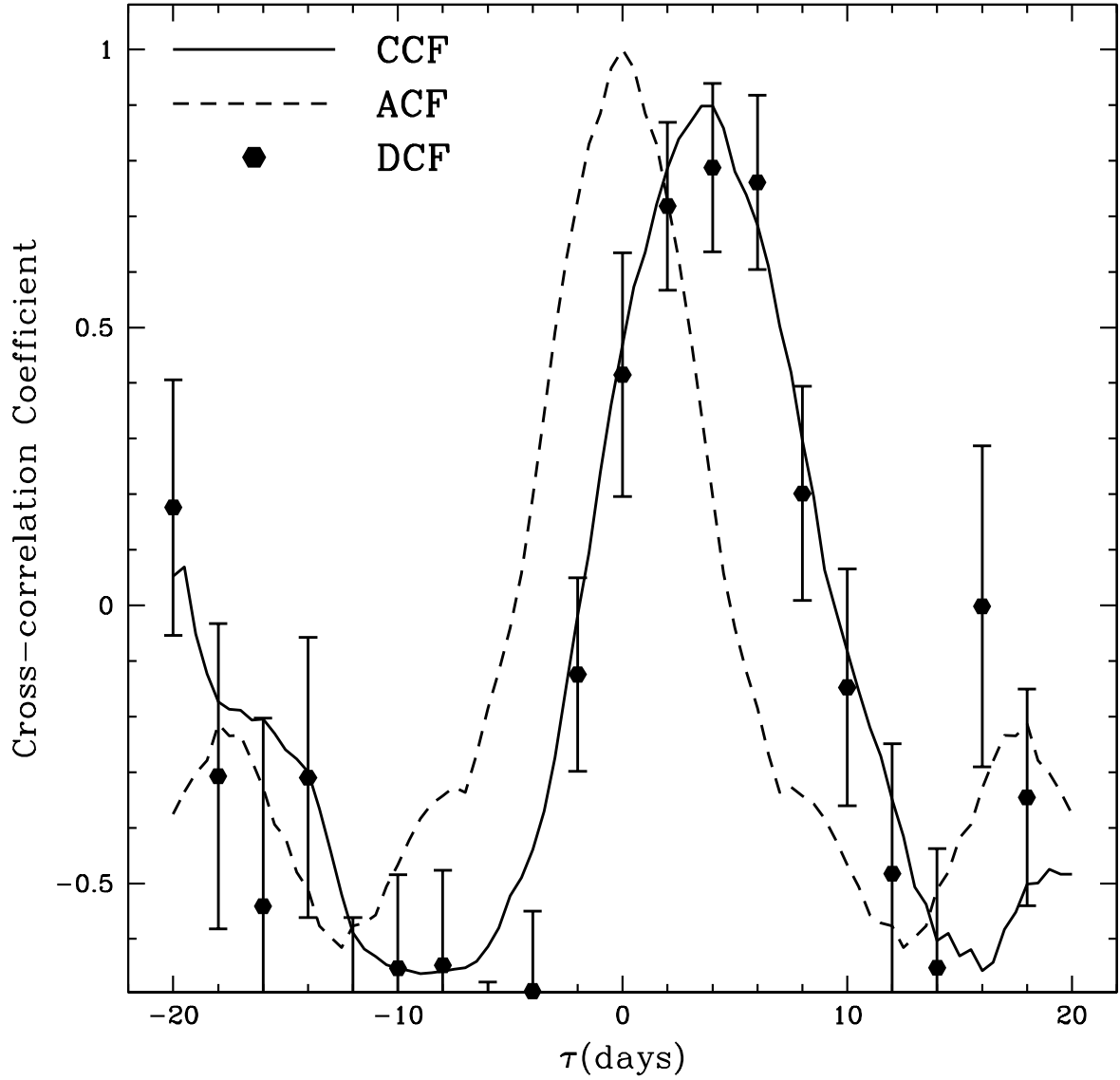}
\caption{Cross correlation function (CCF), discrete correlation function (DCF), and auto correlation function (ACF) from time series analysis of the continuum and H$\beta$ light curves of NGC~4593.}
\end{figure}

\begin{figure}
\figurenum{5}
\epsscale{1}
\plotone{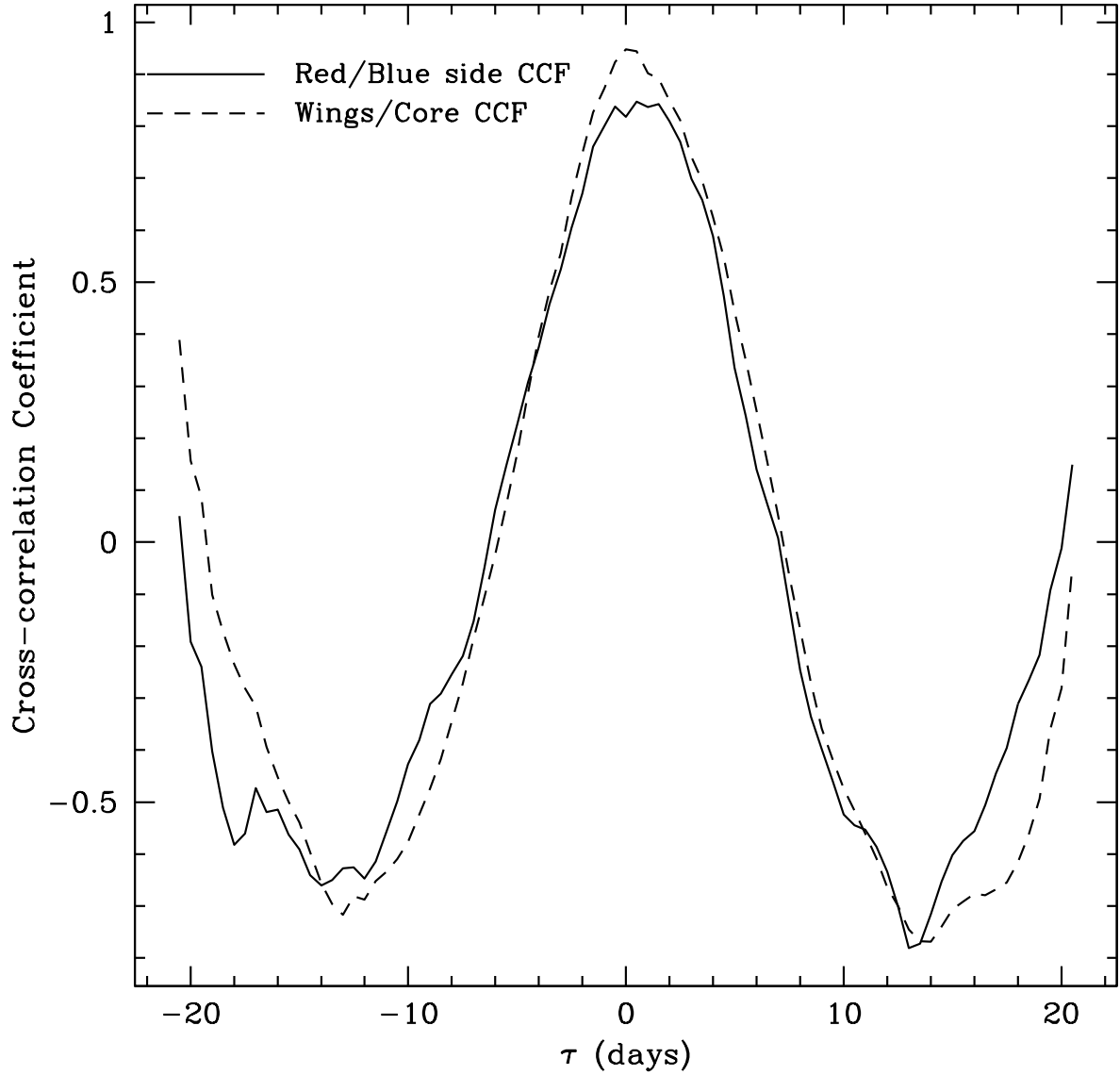}
\caption{Cross correlation functions (CCF) from time series analysis of the red and blue sides of the broad H$\beta$ emission line (solid line) and of the wings and the core of the H$\beta$ emission line (dashed line) of NGC~4593.}
\end{figure}

\end{document}